\documentclass[12pt,a4j]{article} 
\makeatletter
\newdimen{\eqarcolsep}
\usepackage[dvips]{color,graphicx} 
\usepackage{amsmath} 
\usepackage{amsthm} 
\usepackage{bm} 
\usepackage{fancybox} 
\usepackage{ascmac} 
 \usepackage{amssymb}

\makeatletter 
\begin{document}

{\large \bf Semi-regular biorthogonal pairs and generalized Riesz bases}

\ \\
\begin{center}
Hiroshi Inoue\\
\end{center}

\ \\
\begin{abstract} 
In this paper we define the notion of semi-regular biorthogonal pairs what is a generalization of regular biorthogonal pairs in Ref. \cite{hiroshi1} and show that if $(\{ \phi_{n} \} , \{ \psi_{n} \})$ is a semi-regular biorthogonal pair, then $\{ \phi_{n} \}$ and $\{ \psi_{n} \}$ are generalized Riesz bases. This result improves the results of Ref. \cite{h-t, hiroshi1, h-t2} in the regular case.\\
\end{abstract}

\section{Introduction}
Let ${\cal H}$ be a Hilbert space with inner product $( \cdot | \cdot )$, $\bm{e}= \{ e_{n} \}$ an ONB in ${\cal H}$ and $\{ \phi_{n} \}$ a sequence in ${\cal H}$. In Ref. \cite{hiroshi1}, the author has defined an operator $T_{\bm{e}}$ on $D_{\bm{e}} \equiv Span \{ e_{n} \}$ by
\begin{eqnarray}
T_{\bm{e}} \left( \sum_{k=0}^{n} \alpha_{k}e_{k} \right)
= \sum_{k=0}^{n} \alpha_{k} \phi_{k} . \nonumber
\end{eqnarray}
By using this operator $T_{\bm{e}}$, the author has investigated the relationship between a regular biorthogonal pair $( \{ \phi_{n} \} , \{ \psi_{n} \})$ and the notions of Riesz bases and semi-Riesz bases. Here, $( \{ \phi_{n} \} , \{ \psi_{n} \})$ is a pair of Riesz bases if there exists an ONB $\bm{e}= \{ e_{n} \}$ in ${\cal H}$ such that both $T_{\bm{e}}$ and $T_{\bm{e}}^{-1}$ are bounded, and $( \{ \phi_{n} \} , \{ \psi_{n} \})$ is a pair of semi-Riesz bases if there exists an ONB $\bm{e}= \{ e_{n} \}$ in ${\cal H}$ such that either $T_{\bm{e}}$ or $T_{\bm{e}}^{-1}$ are bounded. In this paper we consider the following operators in ${\cal H}$ defined by a sequence $\{ \phi_{n} \}$ in ${\cal H}$ and an ONB $\bm{e}= \{ e_{n} \}$ in ${\cal H}$:
\begin{eqnarray}
T_{\phi, \bm{e}} 
&\equiv& \sum_{k=0}^{\infty} \phi_{k} \otimes \bar{e}_{k} , \nonumber \\
T_{\bm{e},\phi}
&\equiv& \sum_{k=0}^{\infty} e_{k} \otimes \bar{\phi}_{k} , \nonumber
\end{eqnarray}
where the tensor $x \otimes \bar{y}$ of elements $x, \; y$ of ${\cal H}$ is defined by
\begin{eqnarray}
(x \otimes \bar{y} ) \xi = (\xi | y)x, \;\;\; \xi \in {\cal H}. \nonumber
\end{eqnarray} 
This is also denoted by the Dirac notation $|x >< y|$. Here we use the notation $x \otimes \bar{y}$.

In Section 2, we investigate the relationship between the operator $T_{\bm{e}}$ and the operators $T_{\phi, \bm{e}}$ and $T_{\bm{e},\phi}$. The operator $T_{\bm{e},\phi}$ is always closed,
however $D(T_{\phi,\bm{e}}^{\ast})$ is not necessarily dense in ${\cal H}$, equivalently, $T_{\bm{e}}$ and $T_{\phi,\bm{e}}$ are not necessarily closable. Indeed, it is shown that the following statements are equivalent:
\par
\hspace{3mm} (i) $T_{\bm{e}}$ is closable.
\par
\hspace{3mm} (ii) $T_{\phi,\bm{e}}$ is closable.
\par
\hspace{3mm} (iii) $D( T_{\bm{e},\phi} ) = D(\phi) \equiv \left\{ x \in {\cal H} ; \sum_{k=0}^{\infty} | (x | \phi_{k})|^{2} < \infty \right\} $ is dense in ${\cal H}$. \\
If this holds, then
$\bar{T}_{\bm{e}}=\bar{T}_{\phi,\bm{e}}=(T_{\bm{e},\phi})^{\ast}.$ \\
Furthermore we investigate the relationships between the notion of biorthogonal pairs and the operators $T_{\phi,\bm{e}}$, $T_{\bm{e},\phi}$. Indeed, if $D(\phi)$ is dense in ${\cal H}$, then $\bar{T}_{\phi,\bm{e}}$ has an inverse and $\bar{T}_{\phi,\bm{e}}^{-1} \subset T_{\bm{e},\psi}=(T_{\psi,\bm{e}})^{\ast}$. However, $D(\bar{T}_{\phi,\bm{e}}^{-1})$ is not dense in ${\cal H}$ in general. And so we may give the conditions under what $D(\bar{T}_{\phi,\bm{e}}^{-1})$ is dense in ${\cal H}$. In detail, the following statements are equivalent: 
\par
\hspace{3mm} (i) $D_{\phi} \equiv Span \{ \phi_{n} \}$ is dense in ${\cal H}$.
\par
\hspace{3mm} (ii) $T_{\phi,\bm{e}}$ is closable and $\bar{T}_{\phi,\bm{e}}$ has a densely defined inverse.
\par
\hspace{3mm} (iii) $T_{\phi,\bm{e}}^{\ast}(=T_{\bm{e},\phi})$ has a densely defined inverse.\\
If this holds, then $T_{\bm{e},\phi}^{-1}= ( \bar{T}_{\phi,\bm{e}}^{-1})^{\ast}$.\\ 

In Section 3, we first investigate the relationship between semi-regular biorthogonal pairs and generalized Riesz bases. In Definition 2.1 in Ref \cite{h-t}, the author has defined the notion of generalized Riesz bases under the assumption that $D_{\phi}$ and $D_{\psi}$ are dense in ${\cal H}$, and has shown that if $( \{ \phi_{n} \} , \{ \psi_{n} \})$ is a regular biorthogonal pair, then both $\{ \phi_{n} \}$ and $\{ \psi_{n} \}$ are generalized Riesz bases. In this section, we redefine the notion of generalized Riesz bases, that is, $D_{\phi}$ and $D_{\psi}$ are not necessarily dense in ${\cal H}$ and show that if $( \{ \phi_{n} \} , \{ \psi_{n} \})$ is a semi-regular biorthogonal pair, then both $\{ \phi_{n} \}$ and $\{ \psi_{n} \}$ are generalized Riesz bases. This result improves the results of Ref. \cite{h-t, hiroshi1, h-t2}. Furthermore, we have the following results:
\par
\hspace{3mm} (i) If $( \{ \phi_{n} \} , \{ \psi_{n} \})$ is a regular biorthogonal pair, then for any ONB $\bm{e}= \{ e_{n} \}$ in ${\cal H}$. $\bar{T}_{\phi,\bm{e}}$ (resp. $\bar{T}_{\psi,\bm{e}}$) is the minimum among constructing operators of the generalized Riesz basis $\{ \phi_{n} \}$ (resp. $\{ \psi_{n} \}$) and $T_{\bm{e},\psi}^{-1}$ (resp. $T_{\bm{e},\phi}^{-1}$) is the maximum among constructing operator of $\{ \phi_{n} \}$ (resp. $\{ \psi_{n} \}$). Furthermore, any cloesd operator $T$ (resp. $K$) satisfying $\bar{T}_{\phi,\bm{e}} \subset T \subset T_{\bm{e},\psi}^{-1}$ (resp. $\bar{T}_{\psi,\bm{e}} \subset K \subset T_{\bm{e},\phi}^{-1}$) is a constructing operator for $\{ \phi_{n} \}$ (resp. $\{ \psi_{n} \}$).
\par
\hspace{3mm} (ii) If $D(\phi)$ and $D_{\phi}$ are dense in ${\cal H}$, then $\bar{T}_{\phi,\bm{e}}$ (resp. $T_{\bm{e},\phi}^{-1}$) is the minimum (resp. the maximum) among constructing operators of $\{ \phi_{n} \}$ (resp. $\{ \psi_{n} \}$).
\par
\hspace{3mm} (iii) If $D(\psi)$ and $D_{\psi}$ are dense in ${\cal H}$, then $\bar{T}_{\psi,\bm{e}}$ (resp. $T_{\bm{e},\psi}^{-1}$) is the minimum (resp. the maximum) among constructing operators of $\{ \psi_{n} \}$ (resp. $\{ \phi_{n} \}$).\\

We study the physical operators defined by the operators $T_{\phi, \bm{e}}$, $T_{\bm{e},\phi}$, $T_{\psi,\bm{e}}$ and $T_{\bm{e},\psi}$ and an ONB $\bm{e}= \{ e_{n} \}$. If $D(\phi)$ and $D_{\phi}$ are dense in ${\cal H}$, then lowering, raising and number operators $A_{\phi,\bm{e}}$, $B_{\phi,\bm{e}}$ and $N_{\phi,\bm{e}}$ for $\{ \phi_{n} \}$ are defined, respectively, and raising, lowering and number operators $A_{\bm{e},\phi}$, $B_{\bm{e},\phi}$ and $N_{\bm{e},\phi}$ for $\{ \psi_{n} \}$ are defined, respectively. Furthermore, if $D(\psi)$ and $D_{\psi}$ are dense in ${\cal H}$, then lowering, raising and number operators $A_{\psi,\bm{e}}$, $B_{\psi,\bm{e}}$ and $N_{\psi,\bm{e}}$ for $\{ \psi_{n} \}$ are defined, respectively, and raising, lowering and number operators $A_{\bm{e},\psi}$, $B_{\bm{e},\psi}$ and $N_{\bm{e},\psi}$ for $\{ \phi_{n} \}$ are defined, respectively.
These operators connect with ${\it quasi}$-${\it hermitian \; quantum \; mechanics}$, and its relatives. \cite{mostafazadeh, bagarello11, bagarello2013} Many researchers have investigated such operators mathematically. \cite{h-t, h-t2, hiroshi1, b-i-t}

In Section 4, we shall show a method of constructing a semi-regular biorthogonal pair based on the following commutation rule under some assumptions. Here, the commutation rule is that a pair of operators $a$ and $b$ acting on a Hilbert space ${\cal H}$ satisfying
\begin{eqnarray}
ab-ba=I. \nonumber
\end{eqnarray} 
The author has given assumptions to construct the regular biorthogonal pair in Ref. \cite{h-t2}. Indeed, the assumptions in Ref. \cite{h-t2} coincide with the definition of pseudo-bosons as originally given in Ref. \cite{bagarello10}. We shall give some assumptions to construct the semi-regular biorthogonal pair that connect with the definition of pseudo-bosons, and show that by using the results in Section 3 and Ref. \cite{h-t2}, if $D(\phi)$ and $D_{\phi}$ are dense in ${\cal H}$, then we may construct new pseudo-bosonic operators $\{ A_{\phi,\bm{e}}, B_{\phi, \bm{e}}, A_{\bm{e},\phi}, B_{\bm{e},\phi} \}$ and if $D(\psi)$ and $D_{\psi}$ are dense in ${\cal H}$, then we may construct a new pseudo-bosonic operators $\{ A_{\psi,\bm{e}}, B_{\psi, \bm{e}}, A_{\bm{e},\psi}, B_{\bm{e},\psi} , \}$. Furthermore, we investigate the relationship between pseudo-bosonic operators $\{ a,b,a^{\dagger},b^{\dagger} \}$ satisfying some assumptions and the operators $\{ A_{\phi,\bm{e}}, B_{\phi, \bm{e}}, A_{\bm{e},\phi}, B_{\bm{e},\phi} \}$ and $\{ A_{\psi,\bm{e}}, B_{\psi, \bm{e}}, A_{\bm{e},\psi}, B_{\bm{e},\psi} , \}$.

This article is organized as follows. In Section 2, we define new operators $T_{\phi,\bm{e}}$ and $T_{\bm{e},\phi}$ and study the property of these operators. Furthermore, we study the relationship between the operator $T_{\bm{e}}$ and the operators $T_{\phi, \bm{e}}$ and $T_{\bm{e},\phi}$. In Section 3,  we investigate the relationship between semi-regular biorthogonal pairs and generalized Riesz bases and give the physical operators defined by the operators $T_{\phi, \bm{e}}$, $T_{\bm{e},\phi}$, $T_{\psi,\bm{e}}$ and $T_{\bm{e},\psi}$ and an ONB $\bm{e}= \{ e_{n} \}$. In Section 4, we introduce a method of constructing a semi-regular biorthogonal pair based on the pseudo-bosonic operators $\{ a,b,a^{\dagger},b^{\dagger} \}$ under some assumptions and we investigate the relationship between pseudo-bosonic operators satisfying some assumptions and the physical operators $\{ A_{\phi,\bm{e}}, B_{\phi, \bm{e}}, A_{\bm{e},\phi}, B_{\bm{e},\phi} \}$ and $\{ A_{\psi,\bm{e}}, B_{\psi, \bm{e}}, A_{\bm{e},\psi}, B_{\bm{e},\psi} , \}$. In Section 5, we describe future issue with respect to biorthogonal pairs $( \{ \phi_{n} \} , \{ \psi_{n} \})$ and generalized Riesz bases.
\section{Some operators defined by biorthogonal sequences and ONB}
Let ${\cal H}$ be a Hilbert space with inner product $( \cdot | \cdot )$. We consider the following operators in ${\cal H}$ defined by a sequence $\{ \phi_{n} \}$ in a Hilbert space ${\cal H}$ and an ONB $\bm{e}= \{ e_{n} \}$ in ${\cal H}$:
\begin{eqnarray}
T_{\phi, \bm{e}} 
&\equiv& \sum_{k=0}^{\infty} \phi_{k} \otimes \bar{e}_{k} , \nonumber \\
T_{\bm{e},\phi}
&\equiv& \sum_{k=0}^{\infty} e_{k} \otimes \bar{\phi}_{k} . \nonumber
\end{eqnarray}
In Ref. \cite{hiroshi1}, the author have defined an operator $T_{\bm{e}}$ on $D_{\bm{e}} \equiv Span \{ e_{n} \}$ by
\begin{eqnarray}
T_{\bm{e}} \left( \sum_{k=0}^{n} \alpha_{k} e_{k} \right)
= \sum_{k=0}^{n} \alpha_{k} \phi_{k}. \nonumber
\end{eqnarray}
For the operators $T_{\phi,\bm{e}}$, $T_{\bm{e},\phi}$ and $T_{\bm{e}}$ we have the following\\
\par
{\it Lemma 2.1.} {\it The following statements hold.
\par
\hspace{3mm} (1) $T_{\phi, \bm{e}}$ is a densely defined linear operator in ${\cal H}$ such that
\begin{eqnarray}
T_{\phi,\bm{e}} \supset T_{\bm{e}} \;\;\; {\rm and} \;\;\; T_{\phi, \bm{e}} e_{n} =\phi_{n}, \;\;\; n=0,1, \cdots . \nonumber
\end{eqnarray}
\par
\hspace{3mm} (2)
\begin{eqnarray}
D( T_{\bm{e},\phi} ) = D(\phi) \equiv \left\{ x \in {\cal H} ; \sum_{k=0}^{\infty} | (x | \phi_{k})|^{2} < \infty \right\} 
\;\;\; {\rm and} \;\;\; T_{\bm{e}}^{\ast}=T_{\phi, \bm{e}}^{\ast} =T_{\bm{e}, \phi}. \nonumber
\end{eqnarray} }
\par
{\it Proof.} The statements (1) and (2) are easily proved by the definitions of $T_{\phi, \bm{e}}$, $T_{\bm{e}, \phi}$ and $T_{\bm{e}}$.\\\\
By Lemma 2.1, (2), $T_{\bm{e},\phi}$ is closed.
However $D(T_{\phi,\bm{e}}^{\ast})$ is not necessarily dense in ${\cal H}$, equivalently, $T_{\bm{e}}$ and $T_{\phi,\bm{e}}$ are not necessarily closable. Thus we investigate the conditions under what $T_{\phi,\bm{e}}$ is closable.\\
\par
{\it Lemma 2.2.} {\it The following statements are equivalent:
\par
\hspace{3mm} (i) $T_{\bm{e}}$ is closable.
\par
\hspace{3mm} (ii) $T_{\phi,\bm{e}}$ is closable.
\par
\hspace{3mm} (iii) $D(\phi)$ is dense in ${\cal H}$. \\
If this holds, then
\begin{eqnarray}
\bar{T}_{\bm{e}}=\bar{T}_{\phi,\bm{e}}=(T_{\bm{e},\phi})^{\ast}. \nonumber
\end{eqnarray} }
\par
{\it Proof.} This follows from Lemma 2.1, (2).\\\\
Next we study the relationships between the notion of biorthogonal pairs and the operators $T_{\phi,\bm{e}}$, $T_{\bm{e},\phi}$. Then we have the following statements.\\
\par
{\it Lemma 2.3.} {\it Suppose that $( \{ \phi_{n} \} , \{ \psi_{n} \} )$ is a biorthogonal pair such that $D(\phi)$ is dense in ${\cal H}$, then $\bar{T}_{\phi,\bm{e}}$ has an inverse and $\bar{T}_{\phi,\bm{e}}^{-1} \subset T_{\bm{e},\psi}=(T_{\psi,\bm{e}})^{\ast}$.}\\\\
\par
{\it Proof.} By the definitions of $T_{\phi,\bm{e}}$ and $T_{\bm{e},\psi}$, we have
\begin{eqnarray}
T_{\bm{e},\psi}T_{\phi,\bm{e}} e_{n}= T_{\bm{e},\psi} \phi_{n}=e_{n}, \;\;\; n=0,1, \cdots . \nonumber
\end{eqnarray}
Hence we have
\begin{eqnarray}
T_{\bm{e}, \psi}T_{\phi,\bm{e}}=I \;\;\; {\rm on} \;\;\; D_{\bm{e}}. \nonumber
\end{eqnarray}
Thus we have
\begin{eqnarray}
T_{\bm{e},\psi} \bar{T}_{\phi, \bm{e}}= I. \nonumber
\end{eqnarray}
This completes the proof.\\\\\\
In general, $D(\bar{T}_{\phi,\bm{e}}^{-1})$ is not necessarily dense in ${\cal H}$. We investigate the conditions under what $D(\bar{T}_{\phi,\bm{e}}^{-1})$ is dense in ${\cal H}$.\\
\par
{\it Lemma 2.4.} {\it Suppose that $( \{ \phi_{n} \} , \{ \psi_{n} \} )$ is a biorthogonal pair such that $D(\phi)$ is dense in ${\cal H}$. Then the following statements are equivalent: 
\par
\hspace{3mm} (i) $D_{\phi} \equiv Span \{ \phi_{n} \}$ is dense in ${\cal H}$.
\par
\hspace{3mm} (ii) $T_{\phi,\bm{e}}$ is closable and $\bar{T}_{\phi,\bm{e}}$ has a densely defined inverse.
\par
\hspace{3mm} (iii) $T_{\phi,\bm{e}}^{\ast}(=T_{\bm{e},\phi})$ has a densely defined inverse.\\
If this holds, then $T_{\bm{e},\phi}^{-1}= ( \bar{T}_{\phi,\bm{e}}^{-1})^{\ast}$.\\\\}
\par 
{\it Proof.} (i)$\Rightarrow$(ii) Since $D(\phi)$ is dense in ${\cal H}$, by Lemma 2.2 and Lemma 2.3 we have $T_{\phi,\bm{e}}$ is closable and $\bar{T}_{\phi,\bm{e}}$ has an inverse. Furthermore,  since $D(\bar{T}_{\phi,\bm{e}}^{-1})=\bar{T}_{\phi,\bm{e}}D(\bar{T}_{\phi,\bm{e}}) \supset D_{\phi}$ and $D_{\phi}$ is dense in ${\cal H}$, $\bar{T}_{\phi,\bm{e}}^{-1}$ is densely defined.\\
(ii)$\Rightarrow$(iii) By Ref. \cite{hiroshi1}, Lemma 2.2 and Lemma 2.3, we have
\begin{eqnarray}
(\bar{T}_{\phi,\bm{e}}^{-1})^{\ast}
= (\bar{T}_{\phi,\bm{e}}^{\ast})^{-1}
=( T_{\bm{e},\phi})^{-1}. \nonumber
\end{eqnarray}
Hence we have
\begin{eqnarray}
D \left( ( \bar{T}_{\phi,\bm{e}}^{\ast})^{-1} \right)
= T_{\bm{e},\phi} D(T_{\bm{e},\phi}) \supset T_{\bm{e},\phi} D_{\psi} =D_{\bm{e}}. \nonumber
\end{eqnarray}
Thus we have $( T_{\phi,\bm{e}}^{\ast} )^{-1}$ is densely defined.\\
(iii)$\Rightarrow$(i) Take an arbitrary $x \in D_{\phi}^{\perp}$. Then,
\begin{eqnarray}
0=(\phi_{n}|x) =( T_{\phi,\bm{e}} e_{n}|x)=(T_{\bm{e}} e_{n}|x), \;\;\; n=0,1, \cdots . \nonumber
\end{eqnarray}
Hence, by Lemma 2.1, (2) we have
\begin{eqnarray}
x \in D(T_{\bm{e}}^{\ast})=D\left( T_{\phi,\bm{e}}^{\ast} \right) \;\;\; {\rm and} \;\;\; T_{\bm{e}}^{\ast}x=T_{\phi,\bm{e}}^{\ast}x=0. \nonumber
\end{eqnarray}
By (iii), it follows that
\begin{eqnarray}
x= \left( T_{\phi,\bm{e}}^{\ast} \right)^{-1} T_{\phi,\bm{e}}^{\ast} x =0. \nonumber
\end{eqnarray}
Thus, $D_{\phi}$ is dense in ${\cal H}$.
This completes the proof.\\\\\\
Similarly we have the following statements.\\
\par
{\it Lemma 2.5.} {\it Suppose $( \{ \phi_{n} \} , \{ \psi_{n} \} )$ is a biorthogonal pair such that $D(\psi)$ is dense in ${\cal H}$. Then the following statements are equivalent: 
\par
\hspace{3mm} (i) $D_{\psi} \equiv Span \{ \psi_{n} \}$ is dense in ${\cal H}$.
\par
\hspace{3mm} (ii) $T_{\psi,\bm{e}}$ is closable and $\bar{T}_{\psi,\bm{e}}$ has a densely defined inverse.
\par
\hspace{3mm} (iii) $T_{\psi,\bm{e}}^{\ast}(=T_{\bm{e},\psi})$ has a densely defined inverse. \\
If this holds, then $T_{\bm{e},\psi}^{-1}= ( \bar{T}_{\psi,\bm{e}}^{-1})^{\ast}$.\\\\}
\par 
{\it Proof.} This is shown similarly to Lemma 2.4.\\\\
\section{Semi-regular biorthogonal pairs and generalized Riesz bases}
In Ref. \cite{h-t}, the author has defined the notion of generalized Riesz bases. First we redefine the notion of generalized Riesz bases.\\
\par
{\it Definition 3.1.} {\it If there exists a densely defined closed operator $T$ in ${\cal H}$ with a densely defined inverse and there exists an ONB $\bm{e}= \{ e_{n} \}$ in ${\cal H}$ such that
\begin{eqnarray}
\{ e_{n} \} \subset D(T) \cap D \left( (T^{-1})^{\ast} \right) \;\;\; {\rm and} \;\;\; Te_{n} = \phi_{n}, \;\;\; n=0,1, \cdots , \nonumber
\end{eqnarray} 
then a sequence $\{ \phi_{n} \}$ in ${\cal H}$ is called a generalized Riesz basis with a constructing pair $( \bm{e} , T )$.}\\\\
Here, we delete the conditions of Definition 2.1, (ii) and (iii) in Ref. \cite{h-t}, that is, $D_{\phi}$ and $D_{\psi}$ are not necessarily dense in ${\cal H}$. Then we have the following\\
\par
{\it Lemma 3.2.} {\it Let $\{ \phi_{n} \}$ be a generalized Riesz basis. Then, we have the following statements.
\par
\hspace{3mm} (1) $T^{\ast}$ has a densely defined inverse and $(T^{\ast})^{-1}= (T^{-1})^{\ast}$.
\par
\hspace{3mm} (2) $\psi_{n} \equiv (T^{-1})^{\ast} e_{n}$, $n=0,1, \cdots$.
Then, $\{ \phi_{n} \}$ and $\{ \psi_{n} \}$ are biorthogonal and $(T^{-1})^{\ast}$ is a densely defined closed operator in ${\cal H}$ with densely defined inverse $T^{\ast}$. Hence $\{ \psi_{n} \}$ is a generalized Riesz basis with a constructing pair $(\bm{e} , (T^{-1})^{\ast} )$.
\par
\hspace{3mm} (3) $D(\phi) \cap D(\psi)$ is dense in ${\cal H}$.\\}
\par
{\it Proof.} (1) and (2) are easily shown.\\
(3) We first show that 
\begin{equation}
D(T^{\ast}) \subset D(\phi)  \;\;\; {\rm and} \;\;\; R(T)=D(T^{-1}) \subset D(\psi). \tag{2.1}
\end{equation}
Indeed, this follows from
\begin{eqnarray}
\sum_{k=0}^{\infty} |(x| \phi_{k})|^{2}
&=& \sum_{k=0}^{\infty} |(T^{\ast}x |e_{k})|^{2} \nonumber \\
&=& \| T^{\ast}x \|^{2}, \;\;\; x \in D(T^{\ast}) \nonumber
\end{eqnarray}
and
\begin{eqnarray}
\sum_{k=0}^{\infty} |( y| \psi_{k} )|^{2}
&=& \sum_{k=0}^{\infty} |(T^{-1}y |e_{k})|^{2} \nonumber \\
&=& \| T^{-1} y \|^{2}, \;\;\; y \in D(T^{-1}). \nonumber
\end{eqnarray}
Since $D(T^{\ast})$ and $R(T)$ are dense in ${\cal H}$, it follows that $D(\phi)$ and $D(\psi)$ are dense in ${\cal H}$. Next we show that $D(\phi) \cap D(\psi)$ is dense in ${\cal H}$. Take an arbitrary $x \in D(T)$. Let $|T|= \int_{0}^{\infty} \lambda d E_{T}( \lambda)$ be the spectral resolution of the absolute $|T| \equiv (T^{\ast}T)^{\frac{1}{2}}$ of $T$. Then we have $TE_{T}(n) x \in D(T^{\ast}) \cap R(T)$, $n=0,1, \cdots$ and $\lim_{n \rightarrow \infty} TE_{T}(n) x= Tx$. Hence $D(T^{\ast}) \cap R(T)$ is dense in $R(T)$, and since $R(T)$ is dense in ${\cal H}$, it follows from (2.1) that $D(\phi) \cap D(\psi)$ is dense in ${\cal H}$.
This completes the proof.\\\\

In Ref. \cite{hiroshi1}, we have shown that if $( \{ \phi_{n} \} , \{ \psi_{n} \})$ is a regular biorthogonal pair, then both $\{ \phi_{n} \}$ and $\{ \psi_{n} \}$ are generalized Riesz bases. In order to generalize this result, we define the notion of semi-regular biorthogonal pair as follows:\\
\par
{\it Definition 3.3.} {\it A pair $( \{ \phi_{n} \} , \{ \psi_{n} \})$ of biorthogonal sequences in ${\cal H}$ is said to be semi-regular if either $D(\phi)$ and $D_{\phi}$ are dense in ${\cal H}$ or $D(\psi)$ and $D_{\psi}$ are dense in ${\cal H}$.}\\\\\\
We give a concrete example\cite{bagarello13} of semi-regular and non regular biorthogonal bases. Let $\{ e_{n} \}$ be an ONB in ${\cal H}$ and put $\phi_{n}=e_{n}+e_{0}$ and $\psi_{n}=e_{n}$, $n=1,2, \cdots$. Then it is easily shown that $\{ \phi_{n} \}$ and $ \{ \psi_{n} \}$ are biorthogonal bases such that $D_{\phi}$ and $D(\phi)$ are dense in ${\cal H}$, but $D_{\psi}$ is not dense in ${\cal H}$.
We show that if $( \{ \phi_{n} \} , \{ \psi_{n} \})$ is a semi-regular biorthogonal pair, then both $\{ \phi_{n} \}$ and $\{ \psi_{n} \}$ are generalized Riesz bases. In detail, we have the following \\
\par
{\it Theorem 3.4.} {\it Let $\{ \phi_{n} \}$ and $\{ \psi_{n} \}$ be biorthogonal sequences in ${\cal H}$, and let $\bm{e}= \{ e_{n} \}$ be an arbitrary ONB in ${\cal H}$. Then the following statements hold:
\par
\hspace{3mm} (1) Suppose that $( \{ \phi_{n} \} , \{ \psi_{n} \})$ is a regular biorthogonal pair. Then $\{ \phi_{n} \}$ (resp. $\{ \psi_{n} \}$) is a generalized Riesz basis with constructing pairs $(\bm{e}, \bar{T}_{\phi,\bm{e}})$ and $(\bm{e}, T_{\bm{e},\psi}^{-1})$ (resp. $(\bm{e}, \bar{T}_{\psi,\bm{e}})$ and $(\bm{e}, T_{\bm{e},\phi}^{-1})$), and $\bar{T}_{\phi,\bm{e}}$ (resp. $\bar{T}_{\psi,\bm{e}}$) is the minimum among constructing operators of the generalized Riesz basis $\{ \phi_{n} \}$ (resp. $\{ \psi_{n} \}$), and $T_{\bm{e},\psi}^{-1}$ (resp. $T_{\bm{e},\phi}^{-1}$) is the maximal among constructing operators of $\{ \phi_{n} \}$ (resp. $\{ \psi_{n} \}$). Furthermore, any closed operator $T$ (resp. $K$) satisfying $\bar{T}_{\phi,\bm{e}} \subset T \subset T_{\bm{e},\psi}^{-1}$ (resp. $\bar{T}_{\psi,\bm{e}} \subset K \subset T_{\bm{e},\phi}^{-1}$) is a constructing operator for $\{ \phi_{n} \}$ (resp. $\{ \psi_{n} \}$).
\par
\hspace{3mm} (2) Suppose that $D(\phi)$ and $D_{\phi}$ are dense in ${\cal H}$. Then $ \{ \phi_{n} \}$ (resp. $\{ \psi_{n} \}$) is a generalized Riesz basis with a constructing pair $(\bm{e}, \bar{T}_{\phi,\bm{e}})$ (resp. $( \bm{e} , T_{\bm{e}, \phi}^{-1} )$) and the constructing operator $\bar{T}_{\phi ,\bm{e}}$ (resp. $T_{\bm{e},\phi}^{-1}$) is the minimum (resp. the maximum) among constructing operators of $\{ \phi_{n} \}$ (resp. $\{ \psi_{n} \}$).
\par
\hspace{3mm} (3) Suppose that $D(\psi)$ and $D_{\psi}$ are dense in ${\cal H}$. Then $ \{ \psi_{n} \}$ (resp. $\{ \phi_{n} \}$) is a generalized Riesz basis with a constructing pair $(\bm{e}, \bar{T}_{\psi,\bm{e}})$ (resp. $( \bm{e} , T_{\bm{e}, \psi}^{-1} )$) and the constructing operator $\bar{T}_{\psi ,\bm{e}}$ (resp. $T_{\bm{e},\psi}^{-1}$) is the minimum (resp. the maximum) among constructing operators of $\{ \psi_{n} \}$ (resp. $\{ \phi_{n} \}$).\\\\
}
\par
{\it Proof.} Let $\bm{e}= \{ e_{n} \}$ be any ONB in ${\cal H}$.\\
(1) Since $D(\phi)$ is dense in ${\cal H}$, it follows from Lemma 2.3 that $\bar{T}_{\phi, \bm{e}}$ has an inverse. Since $D_{\phi}$ is also dense in ${\cal H}$, it follows from Lemma 2.4 that the inverse $\bar{T}_{\phi,\bm{e}}^{-1}$ of $\bar{T}_{\phi,\bm{e}}$ is densely defined. Furthermore, since $\bar{T}_{\phi,\bm{e}}^{\ast} \psi_{n}=T_{\bm{e},\phi} \psi_{n}=e_{n}$, $n=0,1, \cdots $, we have $ \bm{e} \subset D \left( (\bar{T}_{\phi,\bm{e}}^{\ast})^{-1} \right) =D( T_{\bm{e}, \phi}^{-1})$. Thus $\{ \phi_{n} \}$ is a generalized Riesz basis with a constructing pair $( \bm{e}, \bar{T}_{\phi,\bm{e}} )$, and $\{ \psi_{n} \}$ is a generalized Riesz basis with a constructing pair $(\bm{e} , T_{\bm{e}, \phi}^{-1})$. Similarly, $\{ \psi_{n} \}$ is a generalized Riesz basis with a constructing pair $(\bm{e},\bar{T}_{\psi,\bm{e}})$, and $\{ \phi_{n} \}$ is a generalized Riesz basis with a constructing pair $(\bm{e},T_{\bm{e},\psi}^{-1})$. Hence $\{ \phi_{n} \}$ (resp. $\{ \psi_{n} \}$) is a generalized Riesz basis with constructing pairs $(\bm{e},\bar{T}_{\phi,\bm{e}})$ and $(\bm{e},T_{\bm{e},\psi}^{-1})$ (resp. $(\bm{e},\bar{T}_{\psi,\bm{e}})$ and $(\bm{e},T_{\bm{e},\phi}^{-1})$).

Take an arbitrary constructing operator $T$ of the generalized Riesz basis $\{ \phi_{n} \}$. Since $Te_{n}=\phi_{n}$ and $(T^{-1})^{\ast}e_{n}=\psi_{n}$, $n=0,1, \cdots $, we have $\bar{T}_{\phi,\bm{e}} \subset T$ and $\bar{T}_{\psi,\bm{e}} \subset (T^{-1})^{\ast}$, which implies that $T^{-1} \subset T_{\psi,\bm{e}}^{\ast}=T_{\bm{e},\psi}$. Hence, we have $T \subset T_{\bm{e},\psi}^{-1}$. Thus, $\bar{T}_{\phi,\bm{e}}$ and $T_{\bm{e},\psi}^{-1}$ are the minimum and the maximum among constructing operators of $\{ \phi_{n} \}$, respectively. Furthermore, suppose that $T$ is a closed operator in ${\cal H}$ such that $\bar{T}_{\phi,\bm{e}} \subset T \subset T_{\bm{e},\psi}^{-1}$. Then, since $D(T) \supset D_{\bm{e}}$, $TD(T) \supset T_{\phi,\bm{e}} D_{\bm{e}} = \{ \phi_{n} \}$ and $D((T^{\ast})^{-1}) \supset D(T_{\psi,\bm{e}}) \supset D_{\bm{e}}$, it follows that $T$ is a constructing operator for $\{ \phi_{n} \}$. Similar results for $\{ \psi_{n} \}$ are obtained.\\
The statements (2) and (3) are shown similarly to (1).
This completes the proof. \\
\par
{\it Remark.}  {\it Theorem 3.4 means the following: 
\par
\hspace{3mm} (1) Suppose $D(\phi)$ and $D_{\phi}$ (resp. $D(\psi)$ and $D_{\psi}$) are dense in ${\cal H}$. Even if $D_{\psi}$ (resp. $D_{\phi}$) is not dense in ${\cal H}$, $\{ \psi_{n} \}$ (resp. $\{ \phi_{n} \}$) becomes a generalized Riesz basis.
\par
\hspace{3mm} (2) Suppose that $D(\phi)$ and $D_{\phi}$ are dense in ${\cal H}$, but $D_{\psi}$ is not dense in ${\cal H}$. As shown in Theorem 3.4, $\bar{T}_{\phi,\bm{e}}$ is the minimum among constructing operators of $\{ \phi_{n} \}$, however the maximal constructing operator of $\{ \phi_{n} \}$ does not necessarily exist because $T_{\bm{e},\psi}^{-1}$ is not a constructing operator of $\{ \phi_{n} \}$ different to the case of regular biorthogonal pair. Furthermore, $T_{\bm{e},\phi}^{-1}$ is the maximum among constructing operators of $\{ \psi_{n} \}$, however the minimal constructing operator of $\{ \psi_{n} \}$ does not necessarily exist because $\bar{T}_{\psi,\bm{e}}$ is not a constructing operator of $\{ \psi_{n} \}$. Similar results for the case that $D(\psi)$ and $D_{\psi}$ are dense in ${\cal H}$, but $D_{\phi}$ is not dense in ${\cal H}$ are obtained.}\\\\
By Theorem 3.4, Ref. \cite{h-t} and \cite{h-t2}, we can define the physical operators as follows:\\
\par
(1) Suppose $D(\phi)$ and $D_{\phi}$ are dense in ${\cal H}$. Then, we put
\begin{eqnarray}
A_{\phi,\bm{e}}
&=& \bar{T}_{\phi,\bm{e}} \left( \sum_{k=0}^{\infty} \sqrt{k+1} e_{k} \otimes \bar{e}_{k+1} \right) \bar{T}_{\phi,\bm{e}}^{-1}, \nonumber \\
B_{\phi,\bm{e}}
&=& \bar{T}_{\phi,\bm{e}} \left( \sum_{k=0}^{\infty} \sqrt{k+1} e_{k+1} \otimes \bar{e}_{k} \right) \bar{T}_{\phi,\bm{e}}^{-1}, \nonumber \\
N_{\phi,\bm{e}}
&=& \bar{T}_{\phi,\bm{e}} \left( \sum_{k=0}^{\infty} \sqrt{k+1} e_{k+1} \otimes \bar{e}_{k+1} \right) \bar{T}_{\phi,\bm{e}}^{-1}, \nonumber 
\end{eqnarray}
\begin{eqnarray}
A_{\bm{e},\phi}
&=& T_{\bm{e} , \phi}^{-1} \left( \sum_{k=0}^{\infty} \sqrt{k+1} e_{k+1} \otimes \bar{e}_{k} \right) T_{\bm{e}, \phi}, \nonumber \\
B_{\bm{e},\phi}
&=& T_{\bm{e} , \phi}^{-1} \left( \sum_{k=0}^{\infty} \sqrt{k+1} e_{k} \otimes \bar{e}_{k+1} \right) T_{\bm{e}, \phi}, \nonumber \\
N_{\bm{e},\phi}
&=& T_{\bm{e} , \phi}^{-1} \left( \sum_{k=0}^{\infty} \sqrt{k+1} e_{k+1} \otimes \bar{e}_{k+1} \right) T_{\bm{e}, \phi}. \nonumber 
\end{eqnarray}
\par
(2) Suppose $D(\psi)$ and $D_{\psi}$ are dense in ${\cal H}$. Then, we put
\begin{eqnarray}
A_{\psi,\bm{e}}
&=& \bar{T}_{\psi,\bm{e}} \left( \sum_{k=0}^{\infty} \sqrt{k+1} e_{k} \otimes \bar{e}_{k+1} \right) \bar{T}_{\psi,\bm{e}}^{-1}, \nonumber \\
B_{\psi,\bm{e}}
&=& \bar{T}_{\psi,\bm{e}} \left( \sum_{k=0}^{\infty} \sqrt{k+1} e_{k+1} \otimes \bar{e}_{k} \right) \bar{T}_{\psi,\bm{e}}^{-1}, \nonumber \\
N_{\psi,\bm{e}}
&=& \bar{T}_{\psi,\bm{e}} \left( \sum_{k=0}^{\infty} \sqrt{k+1} e_{k+1} \otimes \bar{e}_{k+1} \right) \bar{T}_{\psi,\bm{e}}^{-1}, \nonumber \\
A_{\bm{e},\psi}
&=& T_{\bm{e} , \psi}^{-1} \left( \sum_{k=0}^{\infty} \sqrt{k+1} e_{k+1} \otimes \bar{e}_{k} \right) T_{\bm{e}, \psi}, \nonumber \\
B_{\bm{e},\psi}
&=& T_{\bm{e} , \psi}^{-1} \left( \sum_{k=0}^{\infty} \sqrt{k+1} e_{k} \otimes \bar{e}_{k+1} \right) T_{\bm{e}, \psi}, \nonumber \\
N_{\bm{e},\psi}
&=& T_{\bm{e} , \psi}^{-1} \left( \sum_{k=0}^{\infty} \sqrt{k+1} e_{k+1} \otimes \bar{e}_{k+1} \right) T_{\bm{e}, \psi}. \nonumber 
\end{eqnarray}
Then we have the following\\
\par
{\it Theorem 3.5.} {\it The following statements hold.
\par
\hspace{3mm} (1) Suppose that $D(\phi)$ and $D_{\phi}$ are dense in ${\cal H}$. Then we have
\begin{eqnarray}
A_{\phi,\bm{e}} \phi_{n}
&=& \left\{
\begin{array}{cl}
0 \;\;\;\;\;\;\;\;\;\; &,n=0, \\
\nonumber \\
\sqrt{n} \phi_{n-1}, \;\;\; &,n=1,2, \cdots,
\end{array}
\right. \nonumber \\
\nonumber \\
B_{\phi,\bm{e}} \phi_{n} &=& \sqrt{n+1} \phi_{n+1}  \;\;\;\; ,n=0,1, \cdots , \nonumber \\
\nonumber \\
N_{\phi,\bm{e}} \phi_{n}
&=& n \phi_{n} , \nonumber \\
\nonumber \\
A_{\bm{e},\phi} \psi_{n}
&=& \sqrt{n+1} \psi_{n+1}  \;\;\;\; ,n=0,1, \cdots , \nonumber \\
\nonumber \\
B_{\bm{e},\phi} \psi_{n}
&=& \left\{
\begin{array}{cl}
0 \;\;\;\;\;\;\;\;\;\; &,n=0, \\
\nonumber \\
\sqrt{n} \psi_{n-1}, \;\;\; &,n=1,2, \cdots,
\end{array}
\right. \nonumber \\
\nonumber \\
N_{\bm{e},\phi} \psi_{n}
&=& n \psi_{n} . \nonumber
\end{eqnarray}
Hence $A_{\phi,\bm{e}}$, $B_{\phi,\bm{e}}$ and $N_{\phi,\bm{e}}$ are lowering, raising and number operators for $\{ \phi_{n} \}$, respectively, and $A_{\bm{e},\phi}$, $B_{\bm{e},\phi}$ and $N_{\bm{e},\phi}$ are raising, lowering and number operators for $\{ \psi_{n} \}$, respectively.
\par
\hspace{3mm} (2) Suppose that $D(\psi)$ and $D_{\psi}$ are dense in ${\cal H}$. Then we have
\begin{eqnarray}
A_{\psi,\bm{e}} \psi_{n}
&=& \left\{
\begin{array}{cl}
0 \;\;\;\;\;\;\;\;\;\; &,n=0, \\
\nonumber \\
\sqrt{n} \psi_{n-1}, \;\;\; &,n=1,2, \cdots,
\end{array}
\right. \nonumber \\
\nonumber \\
B_{\psi,\bm{e}} \psi_{n} &=& \sqrt{n+1} \psi_{n+1}  \;\;\;\; ,n=0,1, \cdots , \nonumber \\
\nonumber \\
N_{\psi,\bm{e}} \psi_{n}
&=& n \psi_{n} , \nonumber \\
\nonumber \\
A_{\bm{e},\psi} \phi_{n}
&=& \sqrt{n+1} \phi_{n+1}  \;\;\;\; ,n=0,1, \cdots , \nonumber \\
\nonumber \\
B_{\bm{e},\psi} \phi_{n}
&=& \left\{
\begin{array}{cl}
0 \;\;\;\;\;\;\;\;\;\; &,n=0, \\
\nonumber \\
\sqrt{n} \phi_{n-1}, \;\;\; &,n=1,2, \cdots,
\end{array}
\right. \nonumber \\
\nonumber \\
N_{\bm{e},\psi} \psi_{n}
&=& n \psi_{n} . \nonumber
\end{eqnarray}
Hence $A_{\psi,\bm{e}}$, $B_{\psi,\bm{e}}$ and $N_{\psi,\bm{e}}$ are lowering, raising and number operators for $\{ \psi_{n} \}$, respectively, and $A_{\bm{e},\psi}$, $B_{\bm{e},\psi}$ and $N_{\bm{e},\psi}$ are raising, lowering and number operators for $\{ \phi_{n} \}$, respectively.\\}
\par
{\it Remark.} {\it 
\par
\hspace{3mm} (i) In case of (1), since
\begin{eqnarray}
A_{\bm{e},\phi}
&=& \left( T_{\phi,\bm{e} }^{-1} \right)^{\ast} \left( \sum_{k=0}^{\infty} \sqrt{k+1} e_{k} \otimes \bar{e}_{k+1} \right)^{\ast} T_{\phi, \bm{e}}^{\ast}, \nonumber \\
B_{\bm{e},\phi}
&=& \left( T_{\phi,\bm{e} }^{-1} \right)^{\ast} \left( \sum_{k=0}^{\infty} \sqrt{k+1} e_{k+1} \otimes \bar{e}_{k} \right)^{\ast} T_{\phi, \bm{e}}^{\ast}, \nonumber \\
N_{\bm{e},\phi}
&=& \left( T_{\phi,\bm{e} }^{-1} \right)^{\ast} \left( \sum_{k=0}^{\infty} \sqrt{k+1} e_{k+1} \otimes \bar{e}_{k+1} \right)^{\ast} T_{\phi, \bm{e}}^{\ast}, \nonumber 
\end{eqnarray}
the author has denoted $A_{\bm{e}}^{\dagger}$, $B_{\bm{e}}^{\dagger}$ and $N_{\bm{e}}^{\dagger}$ in Ref. \cite{h-t2}. 
\par
\hspace{3mm} (ii) Suppose that $D(\phi)$ and $D_{\phi}$ are dense in ${\cal H}$. Then the number operators $N_{\phi,\bm{e}}$ and $N_{\bm{e},\phi}( \equiv N_{\phi,\bm{e}}^{\dagger})$ for $\{ \phi_{n} \}$ and $\{ \psi_{n} \}$, respectively have the relation: $( \bar{T}_{\phi,\bm{e}}^{-1})^{\ast} \bar{T}_{\phi,\bm{e}}^{-1} N_{\phi,\bm{e}}= N_{\phi,\bm{e}}^{\dagger} (\bar{T}_{\phi,\bm{e}}^{-1})^{\ast} \bar{T}_{\phi,\bm{e}}^{-1}$. This is called that $N_{\phi,\bm{e}}$ is a ${\it quasi}$-${\it Hermitian \; operator}$ \cite{s-k2012, s-g-h1992, j-d1961} and positive self-adjoint operator $( \bar{T}_{\phi,\bm{e}}^{-1})^{\ast} \bar{T}_{\phi,\bm{e}}^{-1}$ is often called a metric operator for the ${\it quasi}$-${\it Hermitian \; operator}$ $N_{\phi,\bm{e}}$. Suppose that $D(\psi)$ and $D_{\psi}$ are dense in ${\cal H}$. Then the number operator $N_{\bm{e},\psi}$ is a ${\it quasi}$-${\it Hermitian \; operator}$ for the metric operator $( \bar{T}_{\psi,\bm{e}}^{-1})^{\ast} \bar{T}_{\psi,\bm{e}}^{-1}$. The results on generalized Riesz bases is related to the problem of finding metric operators for quasi-Hermitian operators.}
\section{Semi-regular biorthogonal pairs and Psuedo-bosonic operators}

In this section, we introduce a method of constructing a semi-regular biorthogonal pair based on the following commutation rule under some assumptions. Here, the commutation rule is that a pair of operators $a$ and $b$ acting on a Hilbert space ${\cal H}$ with inner product $( \cdot | \cdot )$ satisfies 
\begin{eqnarray}
ab-ba=I. \nonumber
\end{eqnarray}
In particular, this collapses to the canonical commutation rule (CCR) if $b= a^{\dagger}$. In Ref. \cite{h-t2} the author has shown assumptions to construct the regular biorthogonal pair. Indeed, the assumptions in Ref. \cite{h-t2} coincide with the definition of pseudo-bosons as originally given in Ref. \cite{bagarello10}, where in the recent literature many researchers have investigated. \cite{bagarello13, bagarello10, bagarello11, mostafazadeh, d-t}. In this section, we introduce that some assumptions to construct the semi-regular biorthogonal pair connect with the definition of pseudo-bosons. At first, we construct semi-regular biorthogonal pairs on the above commutation rule. We assume the following statements:\\
\par
{\it Assumption 1.} {\it There exists a non-zero element $\phi_{0}$ of ${\cal H}$ such that 
\par
\hspace{3mm} (i) $a \phi_{0}=0$, 
\par
\hspace{3mm} (ii) $\phi_{0} \in D^{\infty}(b) \equiv \cap_{k=0}^{\infty} D( b^{k})$, 
\par
\hspace{3mm} (iii) $b^{n} \phi_{0} \in D(a)$, $n=0,1, \cdots$.}\\\\
Then, we may define a sequence $\{ \phi_{n} \}$ in ${\cal H}$ by
\begin{eqnarray}
\phi_{n} 
&\equiv& \frac{1}{\sqrt{n!}} \; b^{n} \phi_{0}, \;\;\; n =0,1, \cdots \nonumber \\
&=& \frac{1}{\sqrt{n}} \; b \phi_{n-1}, \;\;\; n =1,2, \cdots . \nonumber 
\end{eqnarray}
Furthermore, we have the following\\
\par
{\it Proposition 4.1.} {\it The following statements hold. 
\par
\hspace{3mm} (1) $b^{n} \phi_{0} \in D(a^{m})$ and 
\begin{eqnarray}
a^{m} b^{n} \phi_{0}
&=& \left\{
\begin{array}{cl}
_{n}P_{m} b^{n-m} \phi_{0} \;\;\;\;\;\;\;\;\;\; &,m\leq n, \\
\nonumber \\
0 \;\;\; &,m > n.
\end{array}
\right. \nonumber 
\end{eqnarray}
\par
\hspace{3mm} (2) $\phi_{n} \in D(N^{m})$ and $N^{m} \phi_{n}=n^{m} \phi_{n}$, $n,m=0,1, \cdots$. In particular, $N\phi_{n}=n \phi_{n}$, $n=0,1, \cdots$.
\par
\hspace{3mm} (3) 
\begin{eqnarray}
a \phi_{n}
&=& \left\{
\begin{array}{cl}
0 \;\;\;\;\;\;\;\;\;\; &,n=0, \\
\nonumber \\
\sqrt{n} \phi_{n-1}, \;\;\; &,n=1,2, \cdots,
\end{array}
\right. \nonumber \\
\nonumber \\
b \phi_{n} &=& \sqrt{n+1} \phi_{n+1}  \;\;\;\; ,n=0,1, \cdots .\nonumber 
\end{eqnarray} }
\par
{\it Proof.} These proofs follow from Ref. \cite{h-t2}.\\
\par
{\it Assumption 2.} {\it There exists a non-zero element $\psi_{0}$ of ${\cal H}$ such that
\par
\hspace{3mm} (i) $b^{\dagger} \psi_{0}=0$, 
\par
\hspace{3mm} (ii) $\psi_{0} \in D^{\infty}(a^{\dagger}) \equiv \cap_{k=0}^{\infty} D( (a^{\dagger})^{k})$, 
\par
\hspace{3mm} (iii) $(a^{\dagger})^{n} \psi_{0} \in D(b^{\dagger})$, $n=0,1, \cdots$.}\\\\ 
Then, we may define a sequence $\{ \psi_{n} \}$ in ${\cal H}$ by
\begin{eqnarray}
\psi_{n} 
&\equiv& \frac{1}{\sqrt{n!}} \; (a^{\dagger})^{n} \psi_{0}, \;\;\; n= 0,1, \cdots \nonumber \\
&=& \frac{1}{\sqrt{n}} \; a^{\dagger} \psi_{n-1}, \;\;\; n =1,2, \cdots . \nonumber 
\end{eqnarray}
And we put an operator $N^{\dagger} \equiv a^{\dagger} b^{\dagger}$. Furthermore we have the following\\
\par
{\it Proposition 4.2.} {\it The following statements hold. 
\par
\hspace{3mm} (1) $(a^{\dagger})^{n} \psi_{0} \in D((b^{\dagger})^{m})$ and 
\begin{eqnarray}
(b^{\dagger})^{m} (a^{\dagger})^{n} \psi_{0}
&=& \left\{
\begin{array}{cl}
_{n}P_{m} (a^{\dagger})^{n-m} \psi_{0} \;\;\;\;\;\;\;\;\;\; &,m\leq n, \\
\nonumber \\
0 \;\;\; &,m > n.
\end{array}
\right. \nonumber 
\end{eqnarray}
\par
\hspace{3mm} (2) $\psi_{n} \in D((N^{\dagger})^{m})$ and $(N^{\dagger})^{m} \psi_{n}=n^{m} \psi_{n}$, $n,m=0,1, \cdots$. In particular, $N^{\dagger} \psi_{n}=n \psi_{n}$, $n=0,1, \cdots$.
\par
\hspace{3mm} (3) 
\begin{eqnarray}
a^{\dagger} \psi_{n} &=& \sqrt{n+1} \psi_{n+1}  \;\;\;\;\;\;\;\;\;\;\; ,n=0,1, \cdots ,\nonumber \\
\nonumber \\
b^{\dagger} \psi_{n}
&=& \left\{
\begin{array}{cl}
0 \;\;\;\;\;\;\;\;\;\; &,n=0, \\
\nonumber \\
\sqrt{n} \psi_{n-1}, \;\;\; &,n=1,2, \cdots .
\end{array}
\right. \nonumber  
\end{eqnarray}}\\
\par
{\it Proof.} These proofs follow from Ref. \cite{h-t2}.\\\\
The above Assumption 1 and Assumption 2 coincide with the assumptions of Ref. \cite{h-t2}. We weaken the assumption of Ref. \cite{h-t2} to the next assumption in order to construct semi-regular biorthogonal pairs. \\
\par
{\it Assumption 3.} {\it \par
\hspace{3mm} Either $D(\phi)$ and $D_{\phi}$ are dense in ${\cal H}$ or $D(\psi)$ and $D_{\psi}$ are dense in ${\cal H}$.} \\\\
Then, if a pair of operators $a$ and $b$ acting on ${\cal H}$ satisfies Assumption 1-3, $( \{ \phi_{n} \} , \{ \psi_{n} \} )$ becomes a semi-regular biorthogonal pair.

By Section 2, Section 3 and Ref. \cite{hiroshi1}, in case of  $D(\phi)$ and $D_{\phi}$ are dense in ${\cal H}$ (resp. $D(\psi)$ and $D_{\psi}$ are dense in ${\cal H}$), $A_{\phi,\bm{e}}$, $B_{\phi,\bm{e}}$ and $N_{\phi,\bm{e}}$ are lowering, raising and number operators for $\{ \phi_{n} \}$, respectively, and $A_{\bm{e},\phi}$, $B_{\bm{e},\phi}$ and $N_{\bm{e},\phi}$ are raising, lowering and number operators for $\{ \psi_{n} \}$, respectively. (resp. $A_{\psi,\bm{e}}$, $B_{\psi,\bm{e}}$ and $N_{\psi,\bm{e}}$ are lowering, raising and number operators for $\{ \psi_{n} \}$, respectively, and $A_{\bm{e},\psi}$, $B_{\bm{e},\psi}$ and $N_{\bm{e},\psi}$ are raising, lowering and number operators for $\{ \phi_{n} \}$, respectively.). And we have
\begin{eqnarray}
A_{\phi,\bm{e}}B_{\phi,\bm{e}} - B_{\phi,\bm{e}}A_{\phi,\bm{e}} \subset I \;\;\; &{\rm  and}& \;\;\;
B_{\bm{e},\phi} A_{\bm{e},\phi} -A_{\bm{e},\phi}B_{\bm{e},\phi}\subset I . \nonumber \\
({\rm resp.} \;\;\; A_{\psi,\bm{e}}B_{\psi,\bm{e}} - B_{\psi,\bm{e}}A_{\psi,\bm{e}} \subset I \;\;\; &{\rm  and}& \;\;\;
B_{\bm{e},\psi} A_{\bm{e},\psi} -A_{\bm{e},\psi}B_{\bm{e},\psi}\subset I .) \nonumber 
\end{eqnarray}
Furthermore, we have the following statements with respect to the operators $A_{\phi,\bm{e}}$, $B_{\phi,\bm{e}}$, $A_{\bm{e},\phi}$ and $B_{\bm{e},\phi}$ (resp. $A_{\psi,\bm{e}}$, $B_{\psi,\bm{e}}$, $A_{\bm{e},\psi}$ and $B_{\bm{e},\psi}$). The proofs are easily shown.\\
\par
{\it Proposition 4.3.} {\it If $D(\phi)$ and $D_{\phi}$ are dense in ${\cal H}$, then the following statements hold.
\par
\hspace{3mm} (1) 
\begin{eqnarray}
\phi_{n}&=& \frac{1}{\sqrt{n !}} B_{\phi, \bm{e}}^{n} \phi_{0}, \;\;\; n=0,1, \cdots , \nonumber \\
\psi_{n}&=& \frac{1}{\sqrt{n!}} A_{\bm{e}, \phi}^{n} \psi_{0}, \;\;\; n=0,1, \cdots . \nonumber
\end{eqnarray}
\par
\hspace{3mm} (2) 
\begin{eqnarray}
A_{\phi, \bm{e}} D_{\phi}
= D_{\phi}, && B_{\phi, \bm{e}}D_{\phi} =D_{\phi} , \nonumber \\
& {\rm and}& \nonumber \\
A_{\bm{e}, \phi} D_{\psi}
= D_{\psi}, &&
B_{\bm{e}, \phi} D_{\psi}
=D_{\psi} . \nonumber
\end{eqnarray} 
}
\par
{\it Proposition 4.4.} {\it If $D(\psi)$ and $D_{\psi}$ are dense in ${\cal H}$, then the following statements hold.
\par
\hspace{3mm} (1) 
\begin{eqnarray}
\psi_{n}&=& \frac{1}{\sqrt{n !}} B_{\psi, \bm{e}}^{n} \psi_{0}, \;\;\; n=0,1, \cdots , \nonumber \\
\phi_{n}&=& \frac{1}{\sqrt{n!}} A_{\bm{e}, \psi}^{n} \phi_{0}, \;\;\; n=0,1, \cdots . \nonumber
\end{eqnarray}
\par
\hspace{3mm} (2) 
\begin{eqnarray}
A_{\psi, \bm{e}} D_{\psi}
= D_{\psi}, && B_{\psi, \bm{e}}D_{\psi} =D_{\psi} , \nonumber \\
& {\rm and}& \nonumber \\
A_{\bm{e}, \psi} D_{\phi}
= D_{\phi}, &&
B_{\bm{e}, \psi} D_{\phi}
=D_{\phi} . \nonumber
\end{eqnarray} 
}
Next we investigate the relationship between pseudo-bosonic operators $\{ a,b,a^{\dagger},b^{\dagger} \}$ satisfying Assumption 1-3 and the operators $A_{\phi,\bm{e}}$, $B_{\phi,\bm{e}}$, $A_{\bm{e},\phi}$ and $B_{\bm{e},\phi}$ ($A_{\psi,\bm{e}}$, $B_{\psi,\bm{e}}$, $A_{\bm{e},\psi}$ and $B_{\bm{e},\psi}$).\\

By Proposition 4.1, Proposition 4.2 and Theorem 3.5 we have the following\\
\par
{\it Lemma 4.5.} {\it The following statements hold.
\par
\hspace{3mm} (1) If $D(\phi)$ and $D_{\phi}$ are dense in ${\cal H}$, then $D(a) \cap D(b) \supset D_{\phi}$,
\begin{eqnarray}
a \lceil_{D_{\phi}} \subset A_{\phi,\bm{e}} \;\;\; {\rm and} \;\;\; b\lceil_{D_{\phi}} \subset B_{\phi,\bm{e}}. \nonumber
\end{eqnarray}
\par
\hspace{3mm} (2) If $D(\psi)$ and $D_{\psi}$ are dense in ${\cal H}$, then $D(a^{\dagger}) \cap D(b^{\dagger}) \supset D_{\psi}$,
\begin{eqnarray}
a^{\dagger} \lceil_{D_{\psi}} \subset B_{\psi,\bm{e}} \;\;\; {\rm and} \;\;\; b^{\dagger} \lceil_{D_{\psi}} \subset A_{\psi,\bm{e}}. \nonumber
\end{eqnarray}
\\}
\par
{\it Proposition 4.6.} {\it The following statements hold.
\par
\hspace{3mm} (1) Suppose that $D(\phi)$ is dense in ${\cal H}$ and $D_{\phi}$ is a core for $\bar{a}$ and $\bar{b}$, then $\bar{a} \subset \bar{A}_{\phi, \bm{e}}$ and $\bar{b} \subset \bar{B}_{\phi,\bm{e}}$. In particular, if $\bar{T}_{\phi,\bm{e}}^{-1}$ is bounded, then $\bar{a} \subset A_{\phi,\bm{e}}=\bar{A}_{\phi, \bm{e}}$ and $\bar{b} \subset B_{\phi,\bm{e}}=\bar{B}_{\phi,\bm{e}}$, and if $\bar{T}_{\phi,\bm{e}}$ is bounded, then $\bar{a} = \bar{A}_{\phi,\bm{e}}$ and $\bar{b} = \bar{B}_{\phi,\bm{e}}$.
\par
\hspace{3mm} (2) Suppose that $D(\psi)$ is dense in ${\cal H}$ and $D_{\psi}$ is a core for $\overline{a^{\dagger}}$ and $\overline{b^{\dagger}}$, then $\overline{a^{\dagger}} \subset \bar{B}_{\psi,\bm{e}}$ and $\overline{b^{\dagger}} \subset \bar{A}_{\psi,\bm{e}}$. In particular, if $\bar{T}_{\psi,\bm{e}}^{-1}$ is bounded, then $\overline{a^{\dagger}} \subset B_{\psi,\bm{e}} = \bar{B}_{\psi,\bm{e}}$ and $\overline{b^{\dagger}} \subset A_{\psi,\bm{e}}=\bar{A}_{\psi,\bm{e}}$, and if $\bar{T}_{\psi, \bm{e}}$ is bounded, then $\overline{a^{\dagger}} = \bar{B}_{\psi,\bm{e}}$ and $\overline{b^{\dagger}} = \bar{A}_{\psi,\bm{e}}$.}\\
\par
{\it Proof.} This is shown similarly to Proposition 2.5 in Ref. \cite{h-t2} by using Lemma 4.5.
\section{Discussions}
As shown in Theorem 3.4, if $(\{ \phi_{n} \} , \{ \psi_{n} \})$ is a semi-regular biorthogonal pair, then $\{ \phi_{n} \}$ and $\{ \psi_{n} \}$ are generalized Riesz bases, and so the physical operators (lowering, raising and number operators) are constructed. In case that $(\{ \phi_{n} \} , \{ \psi_{n} \})$ is not a semi-regular biorthogonal pair, that is, both $D_{\phi}$ and $D_{\psi}$ are not dense in ${\cal H}$, it is meaningful to consider the following question:\\
\par
{\it Question.} {\it Under what conditions is a biorthogonal pair $(\{ \phi_{n} \} , \{ \psi_{n} \})$ a generalized Riesz basis? }\\\\
We have estimated that if a biorthogonal pair $(\{ \phi_{n} \} , \{ \psi_{n} \})$ is a ${\cal D}$-quasi basis\cite{bagarello13, bagarello2013}, then $\{ \phi_{n} \}$ and $\{ \psi_{n} \}$ are generalized Riesz bases, where ${\cal D}$ is a dense subspace in ${\cal H}$ and $(\{ \phi_{n} \} , \{ \psi_{n} \})$ is a ${\cal D}$-quasi basis if 
\begin{eqnarray}
(f,g)
=\sum_{k=0}^{\infty} (f,\psi_{k})(\phi_{k},g)
= \sum_{k=0}^{\infty} (f,\phi_{k})(\psi_{k},g), \nonumber
\end{eqnarray}
for all $f,g \in {\cal D}$

\ \\
Graduate School of Mathematics, Kyushu University, 744 Motooka, Nishi-ku, Fukuoka 819-0395, Japan
\\
h-inoue@math.kyushu-u.ac.jp, 
 \\

\end{document}